\documentclass[10pt,prc,twocolumn,nofootinbib,preprintnumbers]{revtex4-2}
\usepackage{amssymb}
\usepackage{amsmath}
\usepackage{graphicx}
\usepackage[colorlinks=true, pdfstartview=FitV, linkcolor=blue, citecolor=blue, urlcolor=blue]{hyperref}

\begin{document}

\title{Impact of dineutrons on nuclear compositions 
of a core-collapse supernova}

\author{Tatsuya~Matsuki}
\affiliation{Department of Physics, Tokyo University of Science, Tokyo 162-8601, Japan}

\author{Shun~Furusawa}
\affiliation{College of Science and Engineering, Kanto Gakuin University, Kanagawa, Japan}
\affiliation{Interdisciplinary Theoretical and Mathematical Sciences Program (iTHEMS), RIKEN, Wako, Saitama 351-0198, Japan}

\author{Katsuhiko~Suzuki}
\affiliation{Department of Physics, Tokyo University of Science, Tokyo 162-8601, Japan }

\begin{abstract}

We study the nuclear compositions 
in the central region of a core-collapse supernova, assuming the existence of dineutrons ($^2n$) and tetraneutrons ($^4n$).
At 100~ms after core bounce, ${}^2n$ and ${}^4n$ are more abundant than deuterons within radii of approximately 100 and 50~km, respectively. 
Compared to the model ignoring the existence of ${}^2n$ and ${}^4n$, the mass fraction of neutrons up to a radius of 100~km reduces, while the mass fractions of protons, deuterons, and $\rm{{}^4He}$ increase.
Due to the uncertainties in the properties of $^2n$ and $^4n$, we investigate the influence of their binding energies on the nuclear composition. We find the binding energy of $^2n$ has only a modest effect on the overall composition, except for its own mass fraction, while that of $^4n$ has a negligible impact. 

\end{abstract}

\maketitle

\section{Introduction}
Core-collapse supernovae are phenomena that occur when a massive star with a mass of $\gtrsim10\, M_\odot$ dies.
Several simulations have shown that the shock wave generated in core bounce 
is delayed and 
halted by energy loss due to iron decomposition reactions and neutrino emissions. 
Multidimensional effects such as convection
improve the efficiency of this neutrino heating mechanism; moreover, the explosion has been reproduced by these effects in multidimensional simulations~\cite{Shoichi:YAMADA2024pjab.100.015}.

After core bounce and shock propagation, the central region within the shock radius is a hot and dense environment with an excess of neutrons. Light elements such as deuterons and $\rm{{}^4He}$ are abundant in the central region at subsaturation densities
\cite{Furusawa:2022ktu}. In almost all simulations, detailed reactions between light elements and neutrinos are not considered; however, the weak interactions 
may affect neutrino emissions and the shock revival
\cite{Ohnishi:2006mk,Furusawa:2013tta,Fischer:2020krf,Fischer:2015sll,Nasu:2014mta}.

Meanwhile, the existence of multineutron states is under debate~\cite{Ogloblin1989,Marques:2021mqf}. Dineutrons~($^2n$) were theoretically predicted in neutron halo~\cite{Hansen:1987mc}, and some experiments involving neutron-rich nuclei have supported their existence ~\cite{Seth:1991zz,Spyrou:2012zz}.
The dineutron is  
known to be unbound, and its binding energy is believed to be approximately –100~keV in free space, which is  consistent with a negative neutron-neutron scattering length~\cite{Chen:2008zzj,Hammer:2014rba}.
Furthermore, tetraneutrons~($^4n$) have been indirectly observed as both a possible resonant state and a weakly bound state~\cite{Marques:2001wh,Kisamori:2016jie,Duer:2022ehf,Faestermann:2022meh}.

The impact of multineutrons on astrophysics has been a topic of discussion; for example, the role of $^2n$ in nucleosynthesis during the early universe~\cite{Kneller:2003ka,MacDonald_2009} and the potential influence of $^4n$ on the structure of neutron stars~\cite{Ivanytskyi:2019ynz}.
Additionally, Panov and Yudin~\cite{Panov:2019khi} investigated the role of $^2n$ and $^4n$ in high-temperature and subsaturation density environments such as supernova matter and found that ${}^2n$ is present near the central region nearly as much as other light elements such as deuterons.
However, the calculation is based on a spherically symmetric supernova simulation.

In this paper, we focus on the roles of $^2n$ and $^4n$ in the supernova.  Using the results of the supernova simulation with the axial symmetry approximation~\cite{Nagakura:2019evv} to characterize the thermodynamic environment, and assuming the existence of $^2n$ and $^4n$, we calculate the nuclear compositions at 100~ms after core bounce.  We apply the nuclear statistical equilibrium approximation to determine the nuclear compositions, which works well for describing the central region of core-collapse supernovae.

This paper is organized as follows. Section II describes the models used to calculate the nuclear compositions. 
Section III presents the results and Sec. IV gives a summary and discussion of the study.

{\section{Model}
We use the natural unit with $\hbar=1$ and $c=1$ and calculate the nuclear compositions of the supernova matter. The mass density $\rho_B$, the electron ratio $Y_e$, and the temperature $T$ are obtained from the recent supernova simulation through axial symmetry approximation~\cite{Nagakura:2019evv}. Because the temperature is higher than 0.4~MeV, we assume that the nucleons and nuclei are  
in nuclear statistical equilibrium (NSE), 
in which the chemical potential of a nucleus with the neutron number $N$ and the proton number $Z$ is given by
\begin{equation}
\mu_{N,Z} = N\mu_N + Z\mu_Z.
\label{eq:mu}
\end{equation}
Moreover, we assume that the nucleons and nuclei behave as the ideal Boltzmann gases, 
and  their number densities $n_{N, Z}$ are calculated as
\begin{equation}
n_{N,Z} = \kappa g_{N,Z} \left( \frac{M_{N,Z} k_B T}{2\pi} \right)^{3/2}  \\ \exp \left( \frac{\mu_{N,Z} - M_{N,Z}}{k_B T} \right),
\label{eq:boltz}
\end{equation}
where $g_{N,Z}$ is the internal degrees of freedom and $\kappa$ is the factor of the excluded volume effect: $\kappa=1$ for nucleons and $\kappa=1-n_B/n_0$ for nuclei with
the baryon number density $n_B$ and the nuclear saturation density $n_0$~\cite{Furusawa:2022ktu}. 
As for nuclei, a previous study reveals that quantum statistical effects are small~\cite{Furusawa:2020}, supporting the validity of the Boltzmann approximation under the conditions considered in this work.

The mass of the nucleus $M_{N, Z}=Nm_N+Zm_p-B_{N, Z}$ is calculated in terms of the binding energy $B_{N, Z}$ obtained from experiments done on the Earth; 3556 nuclei are considered~\cite{Wang:2021xhn}. 
As for $^2n$ and $^4n$, we follow previous works and adopt
$B_{{}^2n}=-0.066$~MeV~\cite{Panov:2019khi} and $B_{{}^4n}=0.42$~MeV~\cite{Faestermann:2022meh} as the  binding energies of ${ }^2n$ and ${}^4n$, respectively. 
Due to the uncertainty in the binding energies of $^2n$ and $^4n$, we examine the impact on our results by employing different assumed values in Fig.~\ref{fig:BE_di}. 
Moreover, we note that the dineutron may be interpreted as a form of pairing in neutron matter. Thus, more appropriate methods for treating such weakly bound systems may exist, such as the Beth-Uhlenbeck approach and the virial expansion~\cite{Ropke:2014fia,Shen_2011:}, which provide a more systematic description of nucleon correlations and are also applicable to broad resonances.
For simplicity, however, in this study we employ the NSE calculation to evaluate the influence of $^2n$ and $^4n$, as well as their sensitivity to the assumed binding energies.
A more systematic treatment within these advanced frameworks is beyond the scope of the present work but should be addressed in future work.

To determine the nuclear compositions, we solve the following two equations for nucleons and nuclei 
\begin{equation}
\sum_{N, Z} n_{N, Z} (N+Z)=n_B ,
\label{eq:nb}
\end{equation}
\begin{equation}
\sum_{N, Z} n_{N, Z} Z=Y_e n_B .
\label{eq:ne}
\end{equation}
 Equations~\eqref{eq:nb} and \eqref{eq:ne} express the local baryon number conservation and the local charge conservation, respectively.

\section{Results}

We first show the distributions of the density, the temperature, and the electron ratio obtained by the supernova simulation through the axial symmetry approximation at 100 and 200~ms after core bounce \cite{Nagakura:2019evv} in Figs.~\ref{fig:density}, \ref{fig:T}, and \ref{fig:Ye}, respectively.  The central region exhibits higher temperature and density, and is neutron-rich, compared to the outer regions. 

\begin{figure}
    \includegraphics[width=26em]{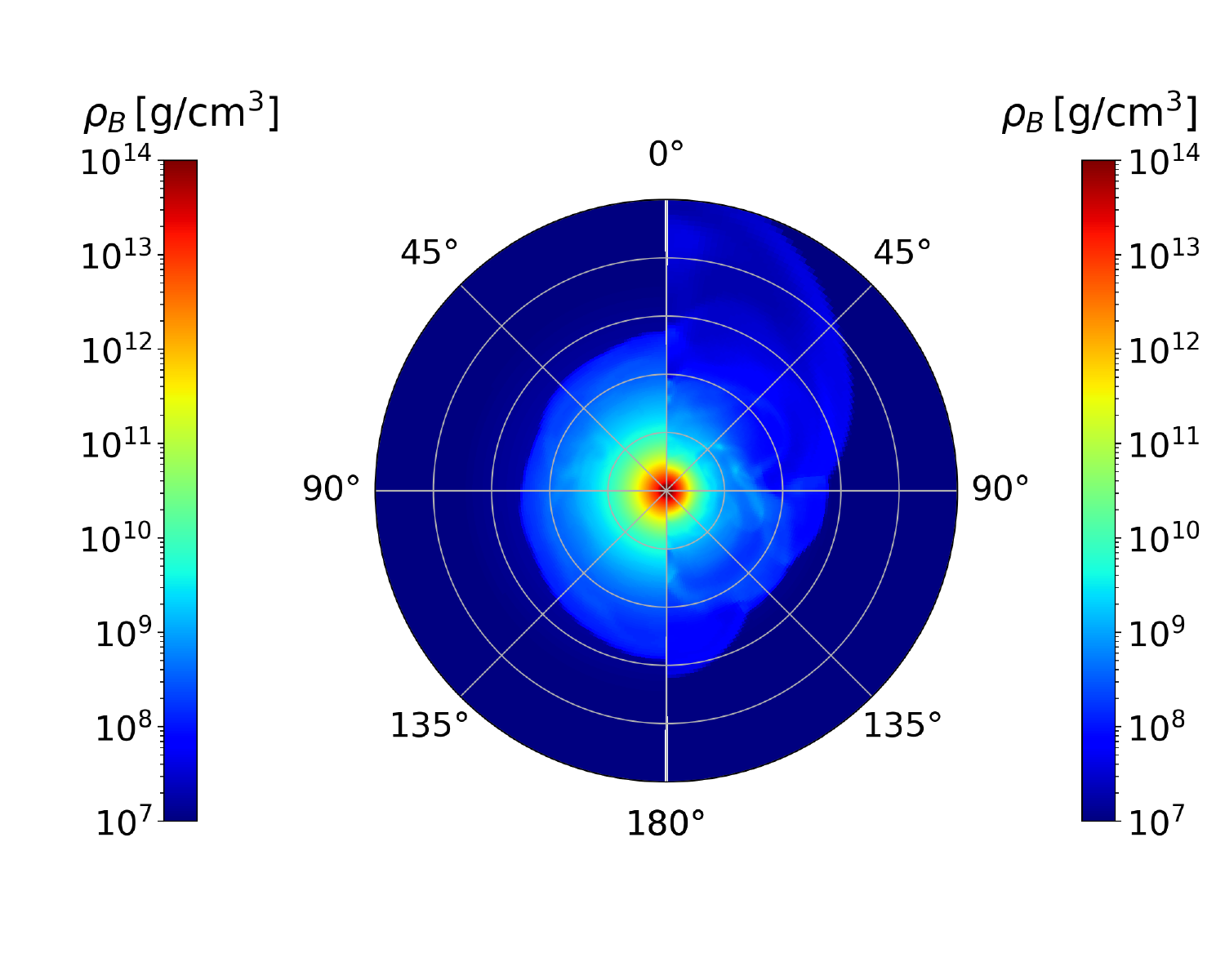}
    \caption{Mass density distributions at 100 and 200~ms after core bounce on the left and right, respectively. The radius is shown up to 500 km, with scale intervals of 100~km.
	}
  \label{fig:density}
\end{figure}

\begin{figure}
    \includegraphics[width=26em]{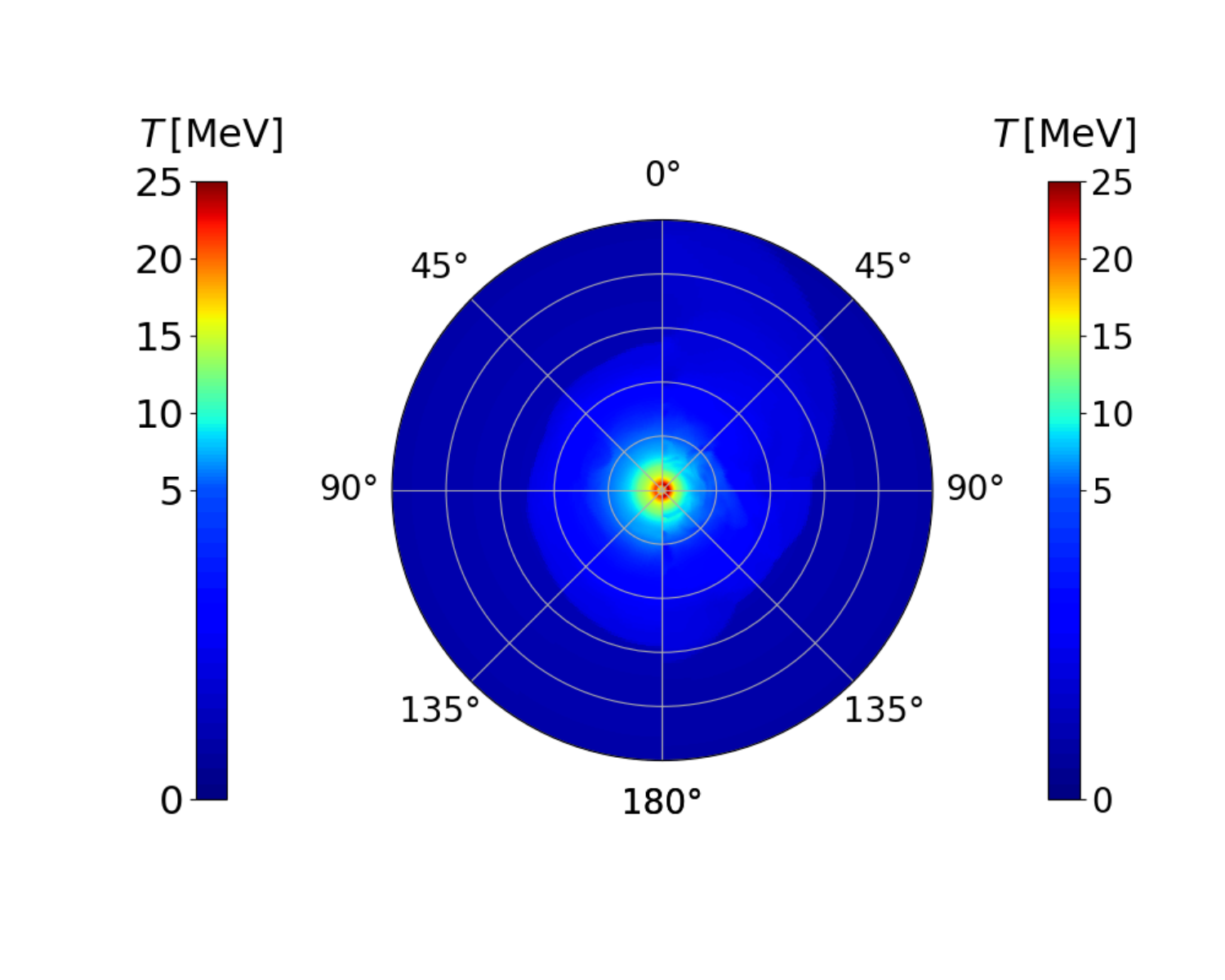}
    \caption{Temperature distributions at 100 and 200~ms after core bounce on the left and right, respectively.
	}
  \label{fig:T}
\end{figure}
\begin{figure}
    \includegraphics[width=26em]{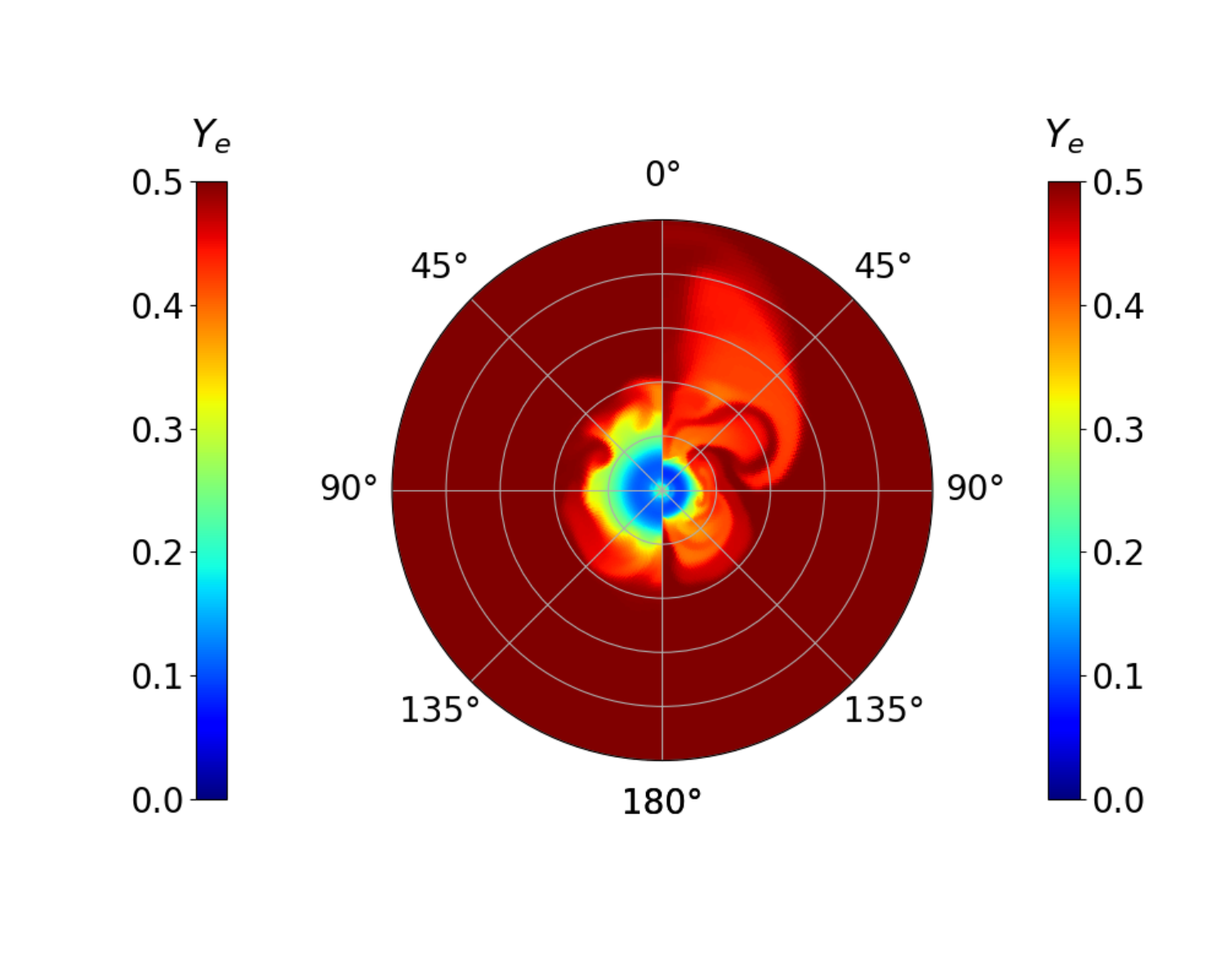}
    \caption{Electron ratio distributions at 100 and 200~ms after core bounce on the left and right, respectively.
	}
  \label{fig:Ye}
\end{figure}

Using the thermodynamical inputs $\rho_B$, $T$ and $Y_e$, we calculate the mass fractions $X_{N,Z} = n_{N,Z} (N+Z) / n_B$ of all nuclei including $^2n$ and $^4n$.  We show a two-dimensional map of the mass fractions of $^2n$ and deuterons at 100 and 200~ms in Figs.~\ref{fig:di-100ms} and \ref{fig:di-200ms}, respectively. Notably, $X_{^2n}$ becomes larger than $X_d$ inside the shock radius, although $X_d$ has been considered to be the largest around protoneutron stars \cite{Furusawa:2022ktu}.
In the following, we focus on the mass fractions of the  $180^\circ$ angle at 100~ms, which is illustrated in  Fig.~\ref{fig:element}.  
In this case, $X_{^4n}$ is large within a radius of approximately 50~km.

Moreover, we add the results for the nuclear composition excluding $^2n$ and $^4n$ in Fig.~\ref{fig:element}, shown by dotted curves.  A significant difference is found within a radius of 100~km.
The mass fraction of neutrons 
decreases due to populations of ${}^2n$ and ${}^4n$. Interestingly, the numbers of the protons, deuterons, and $\rm{{}^4He}$ are found to increase if we include ${}^2n$ and ${}^4n$.

\begin{figure}
    \includegraphics[width=26em]{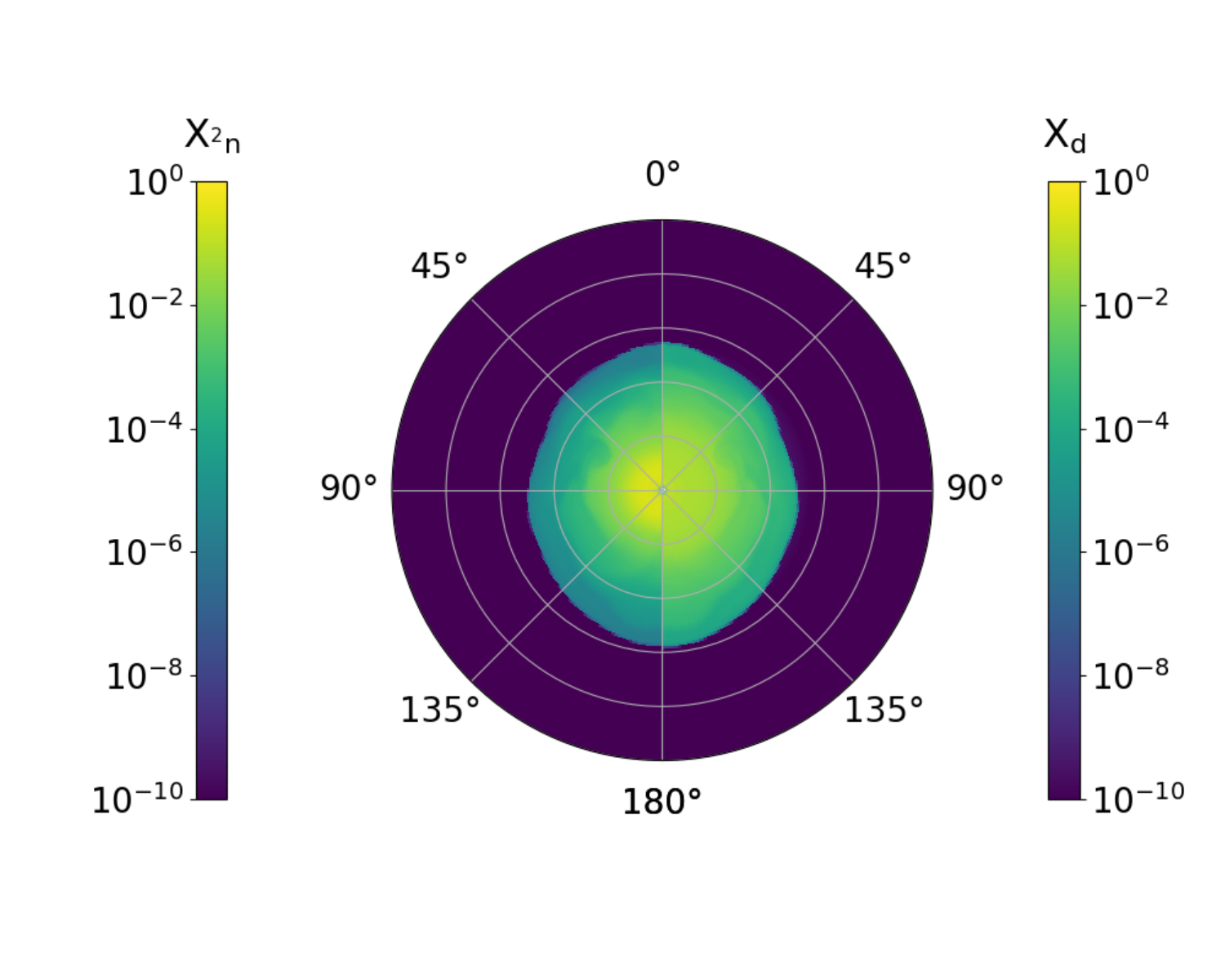}
    \caption{Mass fraction of the dineutron (left) to the deuteron (right) at 100~ms after core bounce.
	}
  \label{fig:di-100ms}
\end{figure}

\begin{figure}
    \includegraphics[width=26em]{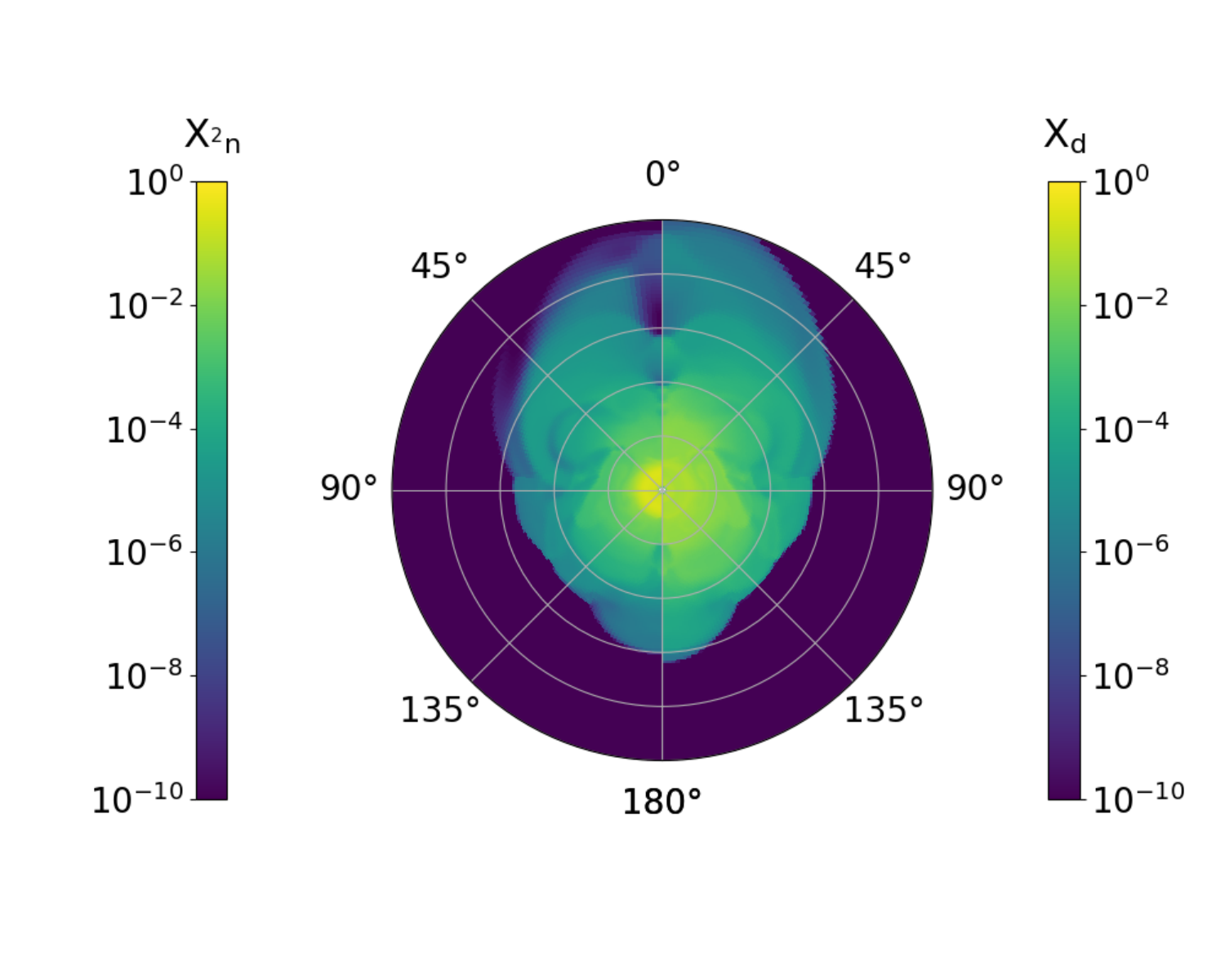}
    \caption{Mass fraction of the dineutron (left) to the deuteron (right) at 200~ms after core bounce.
	}
  \label{fig:di-200ms}
\end{figure}

\begin{figure}
    \includegraphics[width=26em]{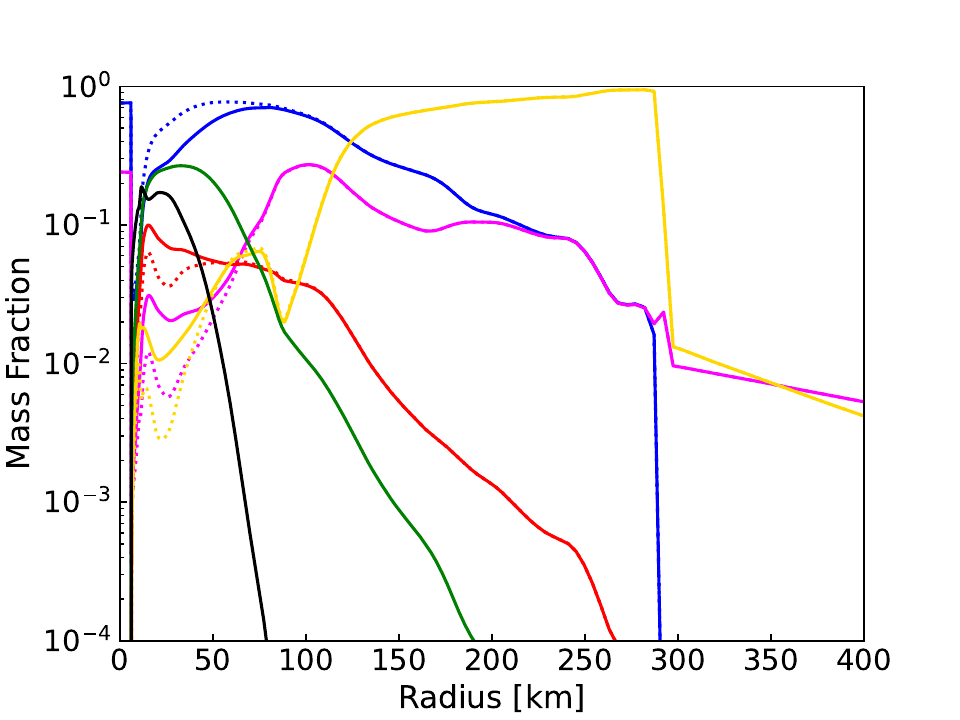}
    \caption{
    Nuclear composition on the $180^\circ$ angle at 100~ms after core bounce. Neutron (blue), proton (magenta), deuteron (red), $\rm{{}^4He}$ (gold),${}^2n$ (green), and ${}^4n$ (black) are displayed. The solid line shows the nuclear composition when the presence of ${}^2n$ and ${}^4n$ is considered, and the dotted line shows one when they are not.
	}
  \label{fig:element}
\end{figure}

The isotopes of hydrogen and helium are shown in Figs.~\ref{fig:H} and \ref{fig:He}, respectively. 
The nuclei of $\rm{{}^3He}$ and $\rm{{}^5He}$ increase in some regions, whereas neutron-rich nuclei such as $\rm{^5H}$ tend to decrease. The decrease of neutrons results in the decrease of the chemical potential of the neutrons. Consequently, the number of neutron-rich nuclei decreases, leading to an increase in the number of protons, deuterons, and $\rm{{}^4He}$ through the total proton conservation
of Eq.~\eqref{eq:ne}.

\begin{figure}
    \includegraphics[width=26em]{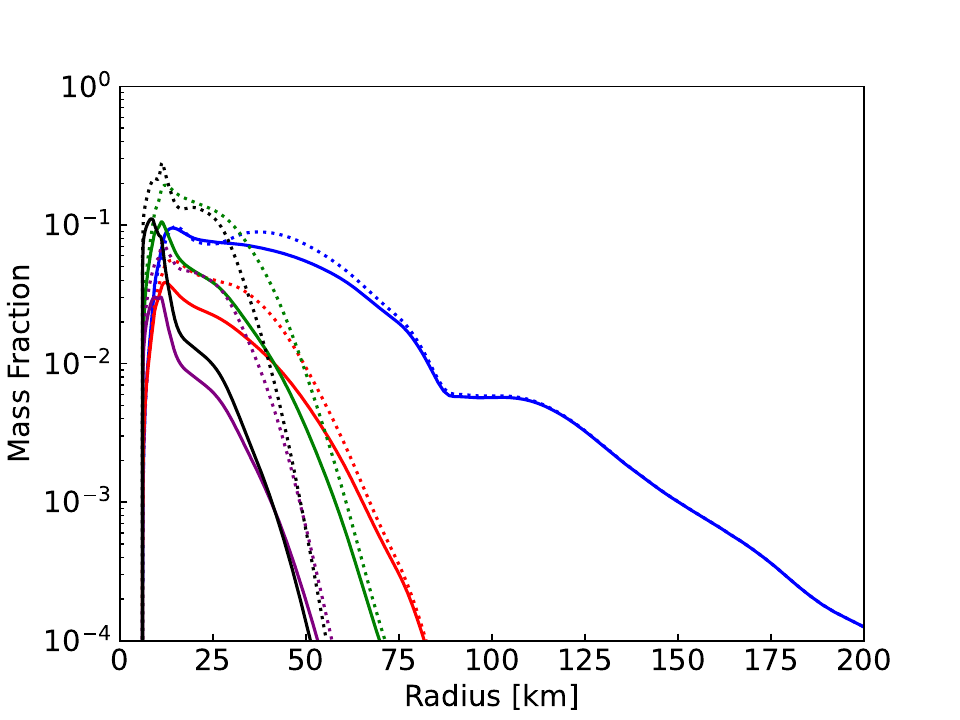}
    \caption{The isotopes of hydrogen within the region up to 200~km. $\rm{{}^3H}$(blue), $\rm{{}^4H}$(red), $\rm{{}^5H}$(green), $\rm{{}^6H}$(purple), and $\rm{{}^7H}$(black) are displayed.
	}
  \label{fig:H}
\end{figure}
\begin{figure}
    \includegraphics[width=26em]{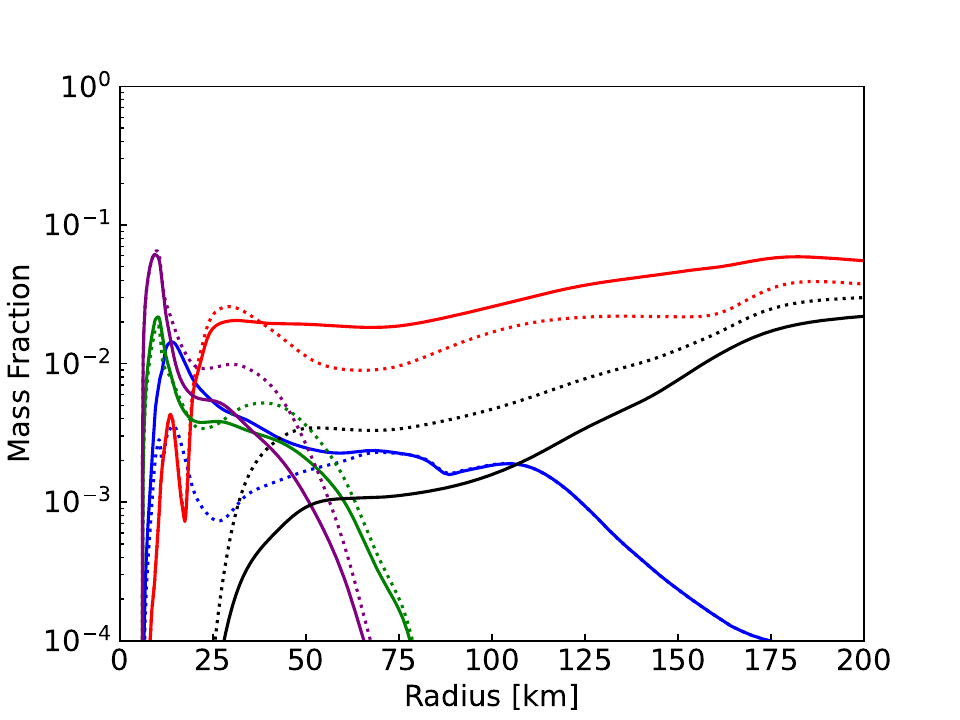}
    \caption{The isotopes of helium within the region up to 200~km. $\rm{{}^3He}$(blue), $\rm{{}^5He}$(red), $\rm{{}^6He}$(green), $\rm{{}^7He}$(purple), and $\rm{{}^8He}$(black) are displayed.
	}
  \label{fig:He}
\end{figure}
For completeness, we calculated the nuclear composition for different values of the binding energy of ${}^2n$ 
and $^4n$
, and the results are presented in Fig.~\ref{fig:BE_di}.
As shown in previous studies \cite{Panov:2019khi}, the binding energy of ${}^2n$ affects its mass fraction; the maximum mass fraction of ${}^2n$ with $B_{{}^2n}=1.0$~MeV
is approximately double that of the model with $B_{{}^2n}=-3.0$~MeV.
The binding energy of $^4n$ similarly affects its mass fraction, but the effect on the overall nuclear composition is negligible.

\begin{figure}
    \includegraphics[width=26em]{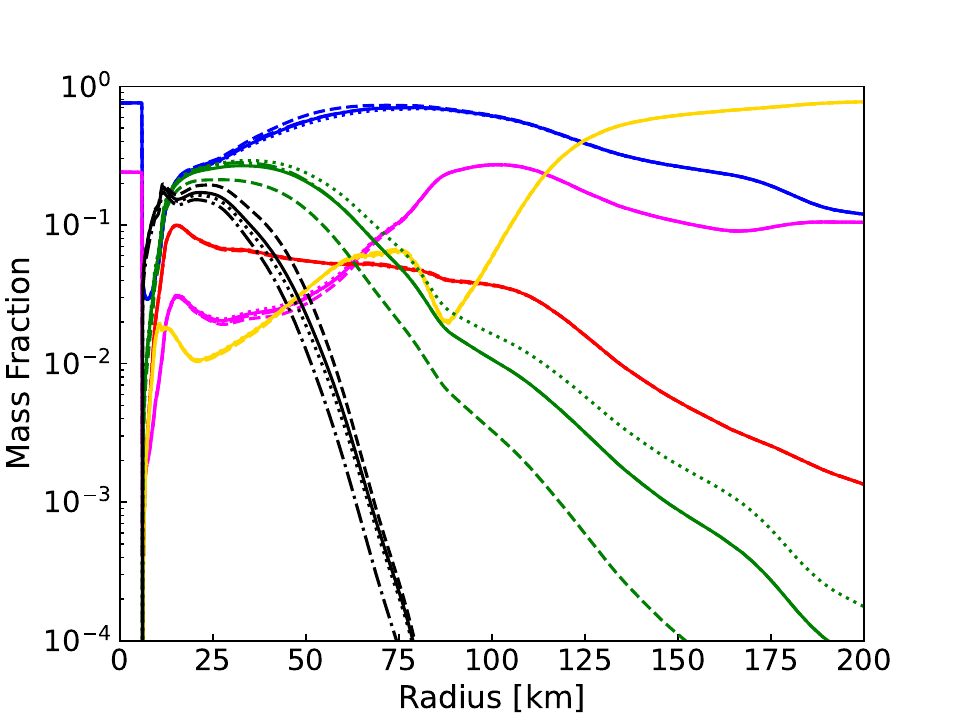}
    \caption{The nuclear compositions for different values of the binding energy of ${}^2n$ and $^4n$. The solid, dashed, and dotted lines represent the nuclear compositions at
    $B_{^2n}=-0.066$, $-3.0$, and 1.0~MeV with $B_{^4n}=0.42$~MeV, respectively. The dash-dotted line represents the nuclear compositions at $B_{^2n}=-0.066$~MeV and $B_{^4n}=–2.37$~MeV~\cite{Duer:2022ehf}.
    The colors in the graph have the same meaning as in Fig.~\ref{fig:element}.
	}
  \label{fig:BE_di}
\end{figure}

\section{Summary and Discussion} 
We have calculated the nuclear compositions 
at 100~ms after core bounce using the results of a two-dimensional simulation \cite{Nagakura:2019evv}, considering the existence of ${}^2n$ and ${}^4n$. Compared to the deuteron, which is reported to be abundant in a previous study \cite{Furusawa:2022ktu}, the mass fractions of ${}^2n$ and ${}^4n$ are larger up to a radius of approximately 100 and 50~km, respectively.
Within a radius of about 100~km, the change in nuclear composition derived from ${}^2n$ and ${}^4n$ is large, where the neutrons decrease and the protons, deuterons, and $\rm{{}^4He}$ increase.
The chemical potential of the neutrons decreases, and consequently, the number of neutron-rich nuclei such as $\rm{}^5H$ decreases, leading to an increase in the number of the protons, deuterons, and $\rm{{}^4He}$.
We also investigate the dependence on the binding energies of multineutron states. They have little impact on the nuclear compositions, except for their own mass fraction.

It is well known that the neutrino reactions with nucleons and light nuclei play essential roles in supernova dynamics~\cite{Sumiyoshi:2008qv}. Owing to changes in nuclear compositions, the existence of $^2n$ and $^4n$ may have significant impacts on the dynamics. A decrease in the number of  free neutrons reduces neutrino absorption reactions, while an increase in free protons and deuterons enhances electron capture reactions, in which protons are converted into neutrons. These changes, occurring in the central region of supernovae, may lead to the earlier neutronization and faster contraction of protoneutron stars. 
For a quantitative study, it is necessary to calculate neutrino reactions with nucleons and light nuclei~\cite{Langanke2003,Nakamura:2000vp} and to construct an equation of state that includes multineutron states. We are currently developing such an equation of state.

However, this study leaves room for further improvement.
First, we adopted the experimental values from the Earth for the binding energy of nuclei; however, those in high-temperature and high-density environments are unknown. The change to a bound state in the environment within the supernova should be clarified and reflected in the calculations. Additionally, the binding energies of ${}^2n$ and ${}^4n$ have not been determined experimentally.
Second, we assumed that the nucleons and nuclei are Boltzmann gases; however, the significant change in the nuclear compositions occurs near the central region of the protoneutron star, which is near the subsaturation density. 
Since the effects of nuclear forces and quantum statistics cannot be ignored, the dependence on the equation of state should be investigated. 
We are currently performing calculations of number densities including their effects based on the relativistic mean-field approximation, and the results are being prepared for publication \cite{Shen:2020sec}.
Finally, we employ the excluded volume effect for light nuclei, but more appropriate approaches may exist, especially for weakly bound states~\cite{Typel2010}. 
The application of these improvements may change the results by a few factors.

\begin{acknowledgments}
We thank K. Mameda and K. Sumiyoshi for productive discussions,
H. Nagakura for providing the simulation data, and T. Uemura for providing the nuclear statistical equilibrium code. S.F. is supported by Grantin-Aid for Scientific Research (Grants No.~19K14723 and No.~24K00632).
\end{acknowledgments}

\bibliography{paper}
\end{document}